\documentclass[eqsecnum,aps,showpacs,twocolumn,epsf]{revtex4} 
\usepackage{graphicx}

\begin{document}
   
 
\title{Mean-field description of collapsing and exploding Bose-Einstein
condensates}
 
\author{Sadhan K. Adhikari}
\address{Instituto de F\'{\i}sica Te\'orica, Universidade Estadual
Paulista, 01.405-900 S\~ao Paulo, S\~ao Paulo, Brazil\\}

\date{\today}
 
 
\begin{abstract}

We perform numerical simulation based on the time-dependent mean-field
Gross-Pitaevskii equation to understand some aspects of a recent
experiment by Donley et al.  on the dynamics of collapsing and exploding
Bose-Einstein condensates of $^{85}$Rb atoms. They   manipulated
the atomic interaction by an external magnetic field via a Feshbach
resonance, thus changing the repulsive condensate into an attractive one
and vice versa. In the actual experiment they changed suddenly the
scattering length of atomic interaction from positive to a large negative
value on a pre-formed condensate in an axially symmetric trap.
Consequently, the condensate collapses and ejects atoms via
explosion. We find that the present mean-field analysis can explain
some aspects of the
 dynamics of the collapsing and exploding Bose-Einstein
condensates.

\end{abstract}

\maketitle
 

\section{Introduction}
 
Recent successful detection \cite{1,1x,ex2} of Bose-Einstein condensates
(BEC) in dilute bosonic atoms employing magnetic trap at ultra-low
temperature has intensified experimental activities on various aspects of
the condensate. On the theoretical front, numerical simulation based on
the time-dependent nonlinear mean-field Gross-Pitaevskii (GP) equation
\cite{8} has
provided a satisfactory account  of some  of these experiments
\cite{11,th1,th1a,th2,th3}.  Since
the detection of BEC for $^7$Li atoms with attractive interaction, one
problem of extreme interest is the dynamical study of the formation and
decay of BEC for attractive atomic interaction \cite{ex2}.

For attractive interaction the condensate is stable for a maximum critical
number $N_{\mbox{cr}}$ of atoms \cite{ex2}.  Measurements for
$N_{\mbox{cr}}$ \cite{ex2,ex3} are in reasonable agreement with the
mean-field analyses
for BEC of $^7$Li in a spherically symmetric trap \cite{11,skxx}, although
there is some
discrepancy for BEC of $^{85}$Rb in an
axially symmetric trap \cite{sk3a,sk1}.  If the number of atoms can
somehow be
increased
beyond
this critical number, due to interatomic attraction the condensate
collapses emitting atoms until the number of atoms is reduced below
$N_{\mbox{cr}}$ and a stable configuration is reached. With a supply of
atoms from an external source the condensate can grow again and thus a
series of collapses can take place, that  was observed experimentally in
the BEC of $^7$Li with attractive interaction \cite{ex2}. Theoretical
mean-field analysis has been able to explain this dynamics
\cite{11,th1,th1a,th2,th3}.

Recently a more challenging experiment has been performed by Donley et al.
\cite{ex4}
at JILA on an attractive condensate of $^{85}$Rb atoms \cite{ex3}
in an axially
symmetric trap, where they
manipulated the interatomic interaction by changing the external magnetic
field exploiting  a nearby Feshbach resonance \cite{fs}. In the vicinity
of a Feshbach resonance the atomic scattering length $a$ can be varied
over a huge range by adjusting the external magnetic field.  Consequently,
they were able to
change suddenly the atomic scattering length by a large amount
for a BEC of $^{85}$Rb atoms \cite{ex3}.
They even changed the sign of the scattering length, thus
transforming a
repulsive condensate into an attractive one. The original experiment on
attractive $^7$Li atoms \cite{ex2} did not use a Feshbach resonance, hence
the atomic
interaction was fixed. This restricts the number of atoms in the  $^7$Li
BEC to a number close to $N_{\mbox{cr}}$ and the collapse is driven by a
stochastic process \cite{ex2,ex4}.
 
 In the experiment conducted at JILA, Donley et al.  changed a
stable pre-formed repulsive condensate of $^{85}$Rb atoms into a highly
explosive and
collapsing attractive condensate and studied the dynamics of collapsing
and exploding condensates \cite{ex4}. The natural scattering length of
$^{85}$Rb atoms is negative (attractive). By exploiting the Feshbach
resonance they made it positive (repulsive) in the initial state,
where the number of atoms, unlike in the experiment with  $^7$Li
\cite{ex2},
could be arbitrarily large. So immediately after the jump
in the scattering length to a large negative value, one has a highly
unstable BEC, where the  number of atoms
could be much larger than $N_{\mbox{cr}}$.  Donley et al.
have provided a quantitative estimate of the explosion of this BEC
by measuring the
number of atoms remaining in the condensate as a function of time until an
equilibrium is reached.  They claim that their experiment reveal many
interesting phenomena that challenge theoretical models.
  Because the phenomenon looks very much like a tiny supernova, or
exploding star, the researchers dubbed it a
          ``Bosenova".  The fundamental physical process underlying the
explosion remains a mystery.

In this paper
we perform a mean-field analysis based on the time-dependent
GP equation to understand some aspects of the above collapse and
explosion of the attractive condensate of $^{85}$Rb atoms in an
axially symmetric trap. To account for the loss of atoms from the
strongly attractive condensate we include an absorptive
nonlinear three-body recombination term in the GP equation.
Three-body recombination leads to the formation of diatomic molecules with
liberation of energy responsible for energetic explosion with
ejection of matter from the BEC. This process  could be
termed ``atomic fusion" in contrast to nuclear fusion in stars.
The
three-body recombination rate we use in numerical  simulation
is in agreement with previous
experimental measurement \cite{k3} and theoretical calculation
\cite{esry}.
The  numerical method, we use,  for the solution of the
time-dependent GP equation with an axially symmetric trap has appeared
elsewhere \cite{sk1,sk2,sk3}. We find that the present mean-field
numerical simulation provides a fair description of some features of the
experiment at JILA \cite{ex4}.

There have been other theoretical studies based on the mean-field GP
equation \cite{th1,th1a,th2,th3} to deal with dynamical collapse including
an
absorptive term to account for the loss of particles. In most cases the
loss mechanism is three-body recombination as in the present
study. However, Duine and Stoof \cite{th1a} propose that the loss arises
due to a new
elastic process.
Instead
of attempting a full numerical solution of the GP equation with axial
symmetry, these investigations used various approximations to study the
time evolution of the condensate or employed a spherically symmetric trap.
Duine and Stoof \cite{th1a} consider the full anisotropic
dynamics, but use a Gaussian approximation for the wave function rather
than an
exact numerical solution. Most of the
other studies employed a spherically
symmetric trap \cite{th1,th2}. However, the investigation
of Ref. \cite{th3} employed an axially symmetric trap to describe some
aspects of the experiment at JILA and
we
comment on this work in Sec. IV.
In the present investigation we consider the complete numerical solution
of the mean-field GP equation for an axially symmetric trap as in the
experiment at JILA. It is realized that an approximate solution as in the
previous
studies can not explain the dynamics of this experiment \cite{th3,ex4}.

In Sec. II we present the theoretical model and the numerical method for
its solution. In Sec, III we present our results that we compare with the
experiment at JILA. Finally, in Secs. IV and V  we present a brief
discussion and concluding remarks.
 
\section{Nonlinear Gross-Pitaevskii Equation}
 
\subsection{Theoretical Model Equations}
 
The time-dependent Bose-Einstein condensate wave
function $\Psi({\bf r};\tau)$ at position ${\bf r}$ and time $\tau $
allowing
for atomic loss
may
be described by the following  mean-field nonlinear GP equation
\cite{8,11}
\begin{eqnarray}\label{a} \biggr[& -& i\hbar\frac{\partial
}{\partial \tau}
-\frac{\hbar^2\nabla^2   }{2m}
+ V({\bf r})
+ gN|\Psi({\bf
r};\tau)|^2-  \frac{i\hbar}{2}
\nonumber \\
& \times & (K_2N|\Psi({\bf r};\tau) |^2
+K_3N^2|\Psi({\bf r};\tau) |^4)
 \biggr]\Psi({\bf r};\tau)=0.
\end{eqnarray}
Here $m$
is
the mass and  $N$ the number of atoms in the
condensate,
 $g=4\pi \hbar^2 a/m $ the strength of interatomic interaction, with
$a$ the atomic scattering length.  A positive $a$ corresponds to a
repulsive interaction and a negative $a$ to an attractive interaction.
The terms $K_2$ and $K_3$ denote two-body
dipolar and three-body recombination loss-rate coefficients, respectively.
There are many ways to account for the loss mechanism \cite{th1,th1a}. It
is
quite impossible to include them all in a self consistent fashion. Here we
simulate the atom  loss via
the most important quintic three-body term  $K_3$ \cite{th1,th2,th3}.
The contribution of the cubic  two-body  loss term
\cite{k3} is
expected to be negligible \cite{th1,th3} compared to the  three-body term
in
the present problem of the  collapsed condensate with large density
and will not be considered here.

The trap potential with cylindrical symmetry may be written as  $  V({\bf
r}) =\frac{1}{2}m \omega ^2(r^2+\lambda^2 z^2)$ where
 $\omega$ is the angular frequency
in the radial direction $r$ and
$\lambda \omega$ that in  the
axial direction $z$. We are using the cylindrical
coordinate system ${\bf r}\equiv (r,\theta,z)$ with $\theta$ the azimuthal
angle.
The normalization condition of the wave
function is
$ \int d{\bf r} |\Psi({\bf r};\tau)|^2 = 1. $

In the absence of angular
momentum the wave function has the form $\Psi({\bf
r};\tau)=\psi(r,z;\tau).$
Now  transforming to
dimensionless variables
defined by $x =\sqrt 2 r/l$,  $y=\sqrt 2 z/l$,   $t=\tau \omega, $
$l\equiv \sqrt {\hbar/(m\omega)}$,
and
\begin{equation}\label{wf}
\phi(x,y;t)\equiv
\frac{ \varphi(x,y;t)}{x} =  \sqrt{\frac{l^3}{\sqrt 8}}\psi(r,z;\tau),
\end{equation}
we get
\begin{eqnarray}\label{d1}
\biggr[-i\frac{\partial
}{\partial t} -\frac{\partial^2}{\partial
x^2}+\frac{1}{x}\frac{\partial}{\partial x} -\frac{\partial^2}{\partial
y^2}
+\frac{1}{4}\left(x^2+\lambda^2 y^2-\frac{4}{x^2}\right) \nonumber \\
+ 8 \sqrt 2 \pi   n\left|\frac {\varphi({x,y};t)}{x}\right|^2
- i\xi n^2\left|\frac {\varphi({x,y};t)}{x}\right|^4
 \biggr]\varphi({ x,y};t)=0,
\end{eqnarray}
where
$ n =   N a /l$
and $\xi=4K_3/(a^2l^4\omega).$
This scaled mean-field equation has the correct $n$ dependence of the
three-body
term so that the same equation can be used to study the decay rate of
different initial and final scattering lengths $a_{\mbox{initial}}$ and
 $a_{\mbox{collapse}}$, respectively, and initial number of atoms $N_0$.
In this study the term $K_3$ will be used for a description of atom loss
in the case of attractive interaction, where the scattering length $a$ is
negative.  From theoretical \cite{ver} and experimental \cite{k3} studies
it has been found that for negative $a,$ $K_3$ increases rapidly as
$|a|^n$, where the theoretical study \cite{ver}
favors $n=2$ for smaller values of $|a|$.
For larger $|a|$, a much larger rate of
increase may take place \cite{esry,ver}.
There are theoretical \cite{esry,ver,k3th}
and experimental \cite{k3ex} estimates  of $K_3$ for $^{87}$Rb,
$^{23}$Na, and $^7$Li away from Feshbach resonance. However,
no thorough and systematic study of the variation of $K_3$ near a Feshbach
resonance
has been performed
\cite{how}.
An  accurate representation of  the variation of $K_3$ of $^{85}$Rb near
the Feshbach resonance
is beyond the scope of this
study and here we represent this variation via a quadratic dependence:
$K_3\sim a^2$. This makes the parameter $\xi$ above a constant for an
 experimental set up with fixed $l$ and $\omega$
and in the present study we  use  a
constant $\xi$.

The normalization condition  of the wave
function becomes
\begin{equation}\label{5} {\cal N}_{\mbox{norm}}\equiv {2\pi} \int_0
^\infty
dx \int _{-\infty}^\infty dy|\varphi(x,y;t)|
^2 x^{-1}=1.  \end{equation}
For $K_3=0,$  ${\cal N}_{\mbox{norm}}=1$, however, in the presence of loss
$K_3 > 0$, ${\cal N}_{\mbox{norm}}
< 1.$ The number of remaining atoms $N$
in the condensate is given by $ N=N_0
{\cal N}_{\mbox{norm}}$,
 where $N_0$ is the initial number.
 
The root mean square (rms) sizes  $x_{\mbox{rms}}$ and  $y_{\mbox{rms}}$
are
defined by
\begin{eqnarray}
x^2_{\mbox{rms}}= {\cal N}_{\mbox{norm}}^{-1} {2\pi} \int_0
^\infty
dx \int _{-\infty}^\infty dy|\varphi(x,y;t)|
^2 x,   \\
y^2_{\mbox{rms}}= {\cal N}_{\mbox{norm}}^{-1} {2\pi} \int_0
^\infty
dx \int _{-\infty}^\infty dy|\varphi(x,y;t)|
^2 y^2x^{-1}.
\end{eqnarray}

\subsection{Numerical Detail}

We solve the GP equation (\ref{d1}) numerically  using a time-iteration
method
elaborated in Refs. \cite{sk1,sk2,sk3,koo}.   The full GP Hamiltonian is
conveniently
broken
into three parts $-$ $H_x$, $H_y$, and $H_n$  $-$ the first containing the
$x$-dependent linear
terms, the second containing the $y$-dependent linear terms and the third
containing the nonlinear terms. The GP equations for the first two  parts
are
defined on a two-dimensional set of grid points $N_x \times N_y$ using the
Crank-Nicholson discretization method. The resultant tridiagonal equations
along $x$ and $y$ directions are solved alternately  by the
Gaussian elimination method along the $x$ and $y$
directions \cite{koo}. The GP equation for the third part do not contain
any space derivative and is solved essentially exactly.
Effectively, each time
iteration of the GP equation is broken up into three parts  $-$
using
$H_x$,
$H_y$ and $H_n$.
For a small time step $\Delta$ the error involved in this
break-up procedure along $x$ and $y$ directions is quadratic in $\Delta$
and hence can be neglected.
For numerical  purpose
we
discretize the GP equation
using time step $\Delta=0.001$ and space step $0.1$ for both
$x$ and $y$ spanning $x$ from 0 to 15 and $y$ from $-30$ to 30. This
domain of space was sufficient to encompass  the whole condensate wave
function even during and after  collapse and explosion. The preparation of
the initial
repulsive wave function is now a routine job and was done by increasing
the nonlinearity $n$ of the GP equation (\ref{d1}) by 0.0001 in each time
step $\Delta$
during time iteration starting with the known harmonic oscillator
solution of Eq. (\ref{d1}) for $n=\xi=0$ \cite{sk1}.

It is now appropriate to calculate the parameters of the present
dimensionless GP equation (\ref{d1}) corresponding to the experiment at
JILA.
We follow the notation and nomenclature of Ref. \cite{ex4}.
Their radial and axial trap frequencies are $\nu_{\mbox{radial}}=17.5$ Hz
and  $ \nu_{\mbox{axial}}=6.8$ Hz,
respectively, leading to $\lambda = 0.389 $. The harmonic oscillator
length $l$ of  $^{85}$Rb atoms for $\omega =2\pi\times 17.5$ Hz and 
$m\approx 79176$
MeV
is
$l=\sqrt{\hbar/(m\omega)}=26070$ \AA. One unit of time $t$ of
Eq. (\ref{d1}) is $1/\omega$ or 0.009095 s.
They prepared a stable $^{85}$Rb
condensate of $N_0= 16000$ atoms with scattering  length
$a_{\mbox{initial}}=7a_0$,
$a_0=0.5292$ \AA, such that the initial $n=2.274$. Then during an
interval of time 0.1 ms the scattering length was ramped to  $a=
a_{\mbox{collapse}}=-30a_0$
such that final $n=-9.744$. The final condensate is strongly attractive
and unstable and undergoes a sequence of collapse and explosion.

The initial value of $n (=2.274)$   was attained after 22740 time steps.
The nonlinearity $n$ is then ramped from 2.274 to
$-9.744$ in 0.1 ms.
As one unit of dimensionless time $t$  is 0.009095 s, 0.1 ms corresponds
to 11  steps of time $\Delta$. In the present simulation, $n$ is  ramped
from 2.274 to $-9.744$ in the GP equation by equal amount in 11
steps.  The absorptive term $\xi$ was set equal
to zero
during above time iteration.
Now the system is prepared for the simulation of the collapse and
explosion.

For the simulation of the collapse and explosion the cubic nonlinear term is
maintained
constant and a nonzero value of $\xi$ is chosen. The time-evolution of
the GP equation is continued as a function of time
$t=\tau_{\mbox{evolve}}$
starting at 0.
The time-evolution is
continued using time step $\Delta =0.001$.  After a small experimentation
it is found that $\xi=2$ fits the experiment at JILA satisfactorily.
Unless otherwise specified,
this value of $\xi$ was used in all   simulations  reported
in this paper for different $a_{\mbox{initial}}$, $a_{\mbox{collapse}}$,
and $N_0$.

It is useful to compare this value of $\xi (=2)$ with the experimental
\cite{k3}
and theoretical \cite{ver} estimates of three-body loss rate of
$^{85}$Rb.  For this we
recall that $K_3=\xi a^2l^4\omega/4.$ Under experimental condition of an
external magnetic field of 250 gauss on $^{85}$Rb \cite{k3} the scattering
length was $a \sim -370a_0$. Consequently, the present value of $\xi (=2)$
corresponds to $K_3 \simeq 9 \times 10^{-25}$ cm$^6$/s for $a\sim
-370a_0$, which is about two
times the experimental rate $K_3=(4.24^{+0.70}_{-0.29}\pm 0.85)\times
10^{-25}$ cm$^6$/s \cite{k3} and about 1.3 times the theoretical rate $K_3
= 6.7\times 10^{-25}$ cm$^6$/s at $a\sim -370 a_0$
\cite{esry}.

\section{Numerical Result}

The numerical simulation using Eq. (\ref{d1}) with a nonzero $\xi$
immediately yields the remaining number of atoms in the condensate
after the jump in scattering length.
The remaining  number of atoms vs. time is plotted in Fig. 1 for
$a_{\mbox{initial}}=7a_0$, $a_{\mbox{collapse}}=-30a_0$, $\xi=2$, and
$N_0=16000$
and compared with the experimental data.
In this figure we also plot the result in this case for $\xi=3$, which
leads to a better agreement with experiment for this specific
case. However, the
use of $\xi=2$
leads to  a more satisfactory  overall agreement with experiment.
Except this single curve in Fig. 1 and the plot in Fig. 4 (a) below, which
are
calculated with $\xi=3$,  all results reported
in this paper are calculated with  $\xi =2$.

\begin{figure}[!ht]
 
\begin{center}
\includegraphics[width=\linewidth]{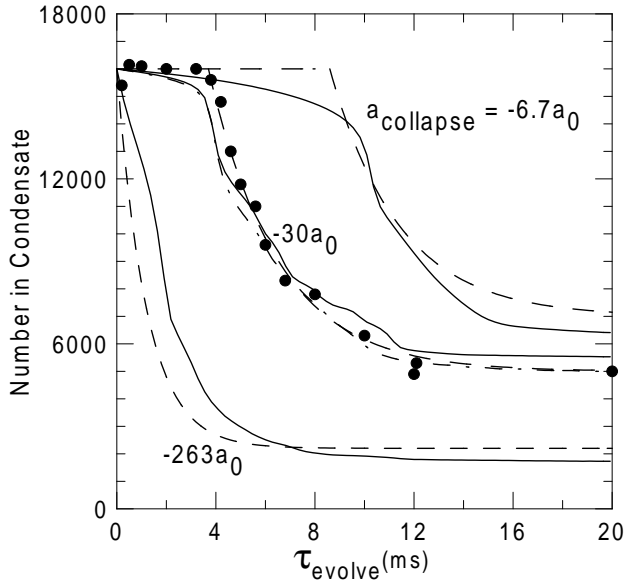}{1.}
\end{center}
 
\caption{ Number of remaining atoms in the condensate of 16000 $^{85}$Rb atoms
after ramping the scattering length from $a_{\mbox{initial}}=7a_0$ to
$a_{\mbox{collapse}}= $ $-6.7a_0,$ and $ -30a_0, $ $-263a_0$ in 0.1 ms as
a function of evolution time $\tau_{\mbox{evolve}}$ in ms. Solid circle:
experiment for $a_{\mbox{collapse}} =-30a_0$ \cite{ex4}; full line: theory
($\xi=2$); dash-dot line: theory ($\xi=3$, $a_{\mbox{collapse}} =-30a_0$);
dashed line: average over preliminary, unanalized data using Eq.
(\ref{dec}) \cite{don1}.}

\end{figure}

In the experiment at JILA \cite{ex4} it was observed that the strongly
attractive  condensate after preparation remains stable with a constant
number of atoms
for an interval of
time $t_{\mbox{collapse}}$, called collapse time.
 This behavior is physically expected. Immediately after
the jump in scattering length from $7a_0$ to $-30a_0$, the attractive
condensate shrinks in size during
$t_{\mbox{collapse}}$,   until the central
density increases to a maximum. Then the absorptive three-body term takes
full control to initiate the 
explosion. Consequently, the number of atoms remains constant
for
$\tau_{\mbox{evolve}}<t_{\mbox{collapse}}$.
The present result (full line) also shows a similar
behavior. However, in this simulation  the absorptive term is
operative from $\tau_{\mbox{evolve}}=0$ and the atom number decreases
right from beginning, albeit at a much smaller rate for
$\tau_{\mbox{evolve}}<t_{\mbox{collapse}}$.

Donley et al.  repeated their experiment with different values of
$a_{\mbox{initial}}$, $a_{\mbox{collapse}}$, and $N_0$
\cite{ex4}.
For $a_{\mbox{initial}}=7a_0$
we  repeated our calculation with the following values of
final scattering length: $a_{\mbox{collapse}}=-263a_0$ and $-6.7a_0$.
These results are also plotted in Fig. 1 and agree with the
unpublished, preliminary unanalyzed data
 \cite{don1}.
The initial delay
 $t_{\mbox{collapse}}$ in starting the explosion is large for small
$|a_{\mbox{collapse}}| $
 as we see in Fig. 1.
Similar effect was observed in the experiment for an initial
condensate of 6000 atoms as shown in their Fig. 2 \cite{ex4}.
After a sequence of collapse and explosion,
Donley et al. observed a ``remnant" condensate of
$N_{\mbox{remnant}}$ atoms at large times containing a certain
constant fraction of
the initial $N_0$ atoms. Figure 1 shows such a behavior.

\begin{figure}[!ht]
 
\begin{center}
\includegraphics[width=\linewidth]{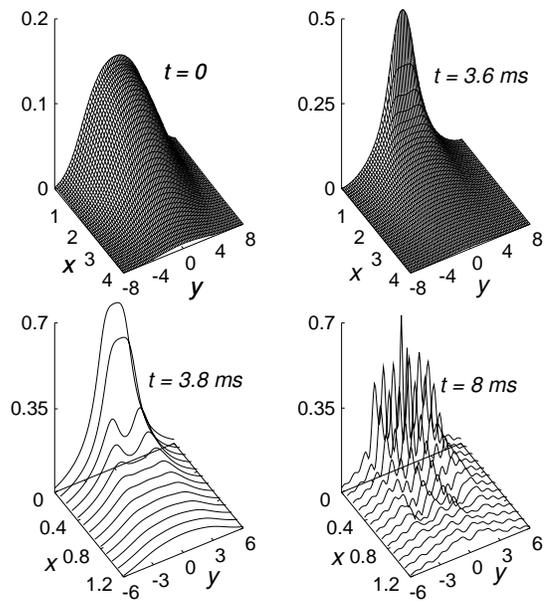}
\end{center}
 
\caption{The central part of the dimensionless wave function $|\phi(x,y)|\equiv
|\varphi(x,y)/x|$ of the condensate on $0.1\times 0.1$ grid for $\xi=2$
after the jump in the scattering length of a BEC of 16000 $^{85}$Rb atoms
from $a_{\mbox{initial}}=7a_0$ to $a_{\mbox{collapse}}=-30a_0$ at times
$\tau_{\mbox{evolve}} =0$, 3.6 ms, 3.8 ms, and 8 ms. The quantities $x$
and $y$ are expressed in units of $l/\sqrt 2$, where $l=26070$ \AA.}

\end{figure}

The above evolution of the condensate after the jump in scattering length
to $-30a_0$ from $7a_0$ for $N_0=16000$ can be understood from a study of
the wave function and we display the central part of the wave function
in
Fig. 2 for $\tau_{\mbox{evolve}}=0, 3.6, 3.8, $ and $8$ ms. The wave
function immediately after jump at time $\tau _{\mbox{evolve}}=0$ is
essentially the same as that before the jump at $-0.1$ ms. There is not
enough time for the wave function to get modified at
$\tau_{\mbox{evolve}}=0$. From Fig. 2 we find that at 3.6 ms the wave
function is only slightly narrower than at 0 ms but still smooth and has
not yet collapsed sufficiently. As $\tau _{\mbox{evolve}}$ increases, the
wave function contracts further and the explosion starts. At 3.8 ms some
spikes (irregularities) have appeared in the wave function showing the
beginning of the explosion and loss. From the study of the wave functions we
find that the explosion start at $\tau
_{\mbox{evolve}}=t_{\mbox{collapse}}\simeq 3.7 $ ms in agreement with the
experiment at JILA.
We also find that at 3.7 ms before the loss began
the bulk BEC did not contract dramatically as also observed in the
experiment. In the numerical simulation for this case we find that at
$\tau _{\mbox{evolve}}=0, x_{\mbox{rms}} =2.98 \mu$m and $y_{\mbox{rms}}
=4.21 \mu$m and at $\tau _{\mbox{evolve}}=3.7$ ms, $x_{\mbox{rms}} =2.53
\mu$m and $y_{\mbox{rms}} =4.10 \mu$m. From Fig. 2 we see that at 8 ms the
wave function is very spiky corresponding to the violent ongoing
explosion.

\begin{figure}[!h]
 
\begin{center}
\includegraphics[width=\linewidth]{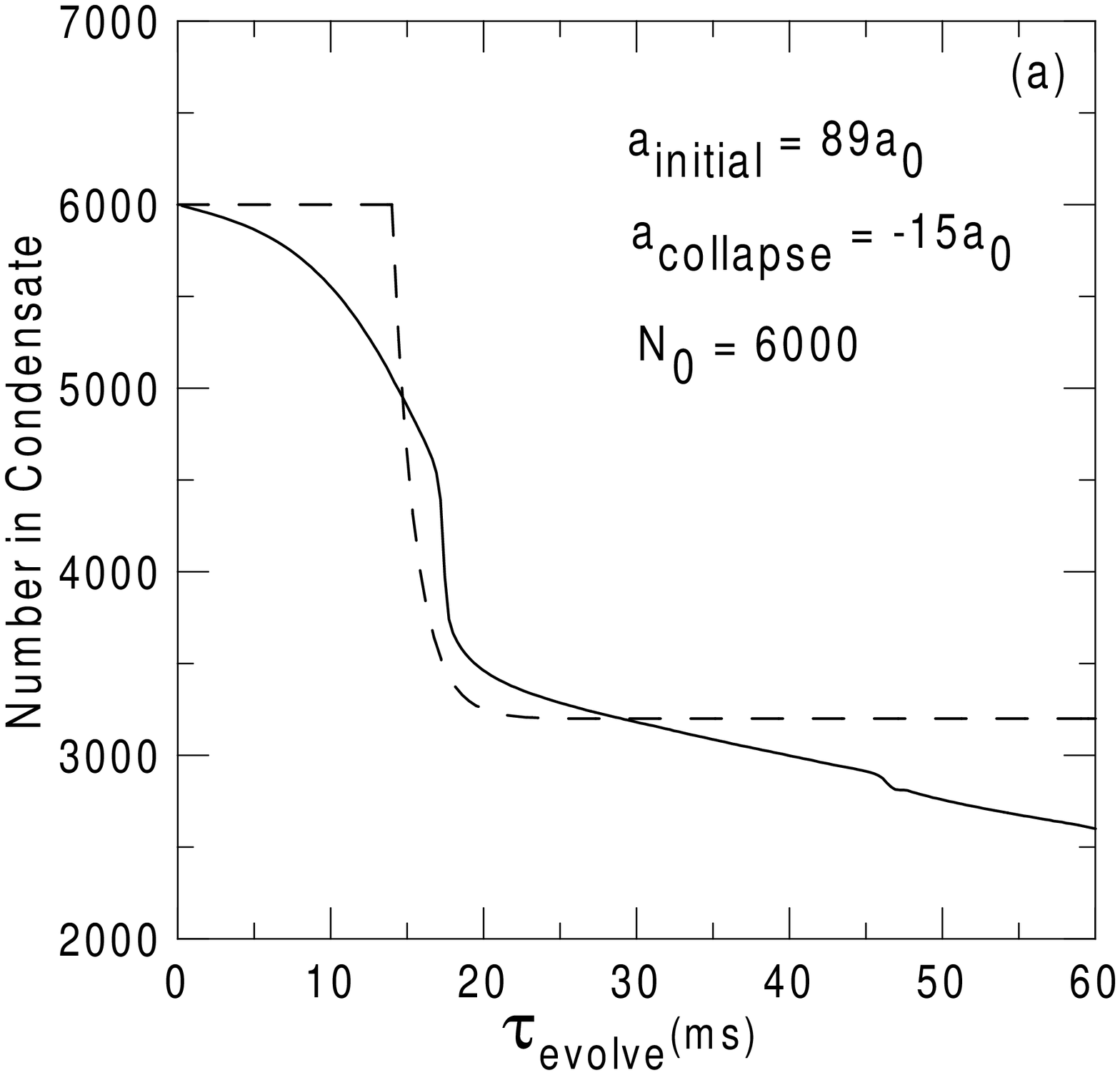}
\end{center}

\end{figure}

\begin{figure}[!h]
 
\begin{center}
\includegraphics[width=\linewidth]{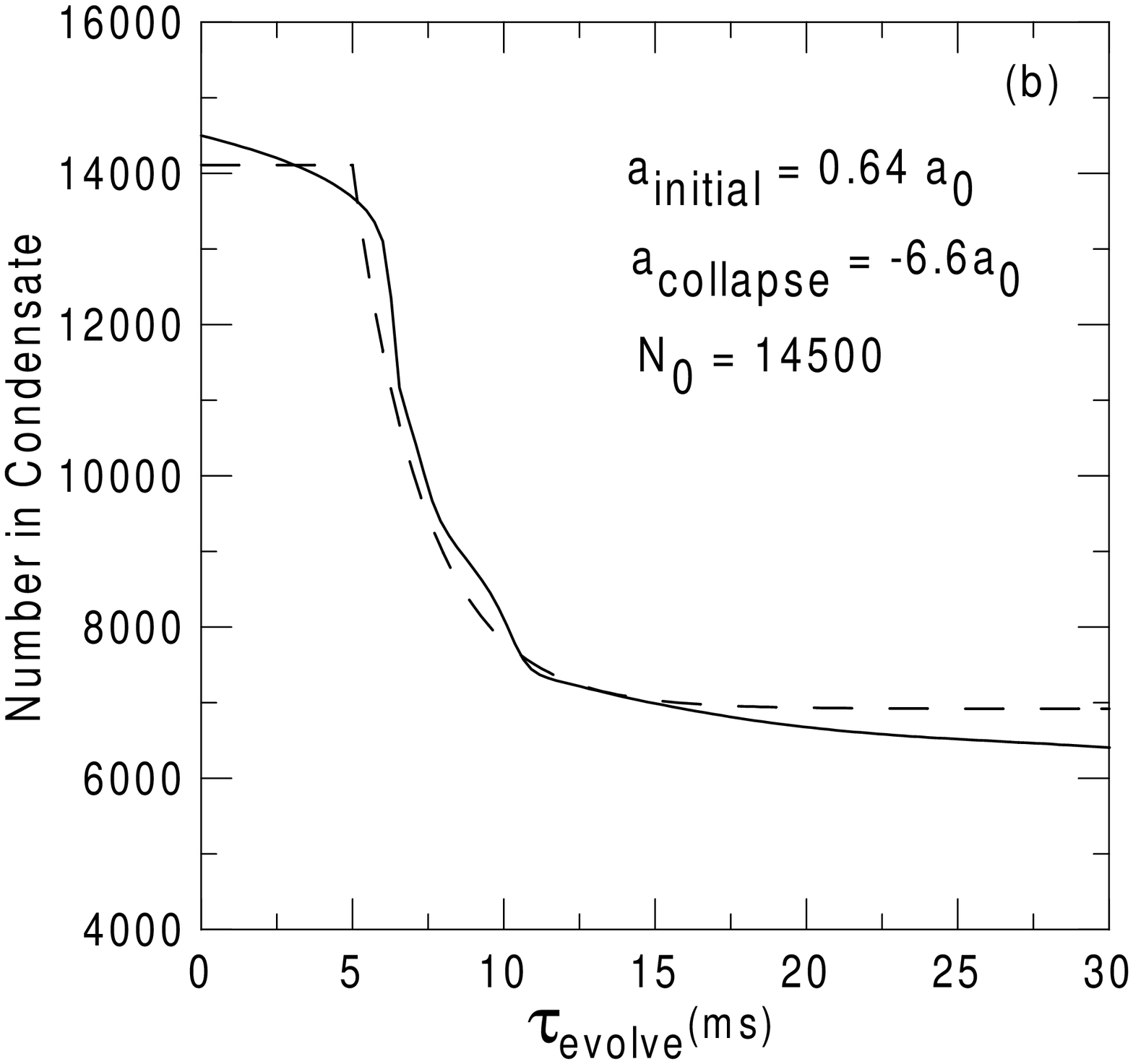}
\end{center}
 
\end{figure}

\begin{figure}
 
\begin{center}
\includegraphics[width=\linewidth]{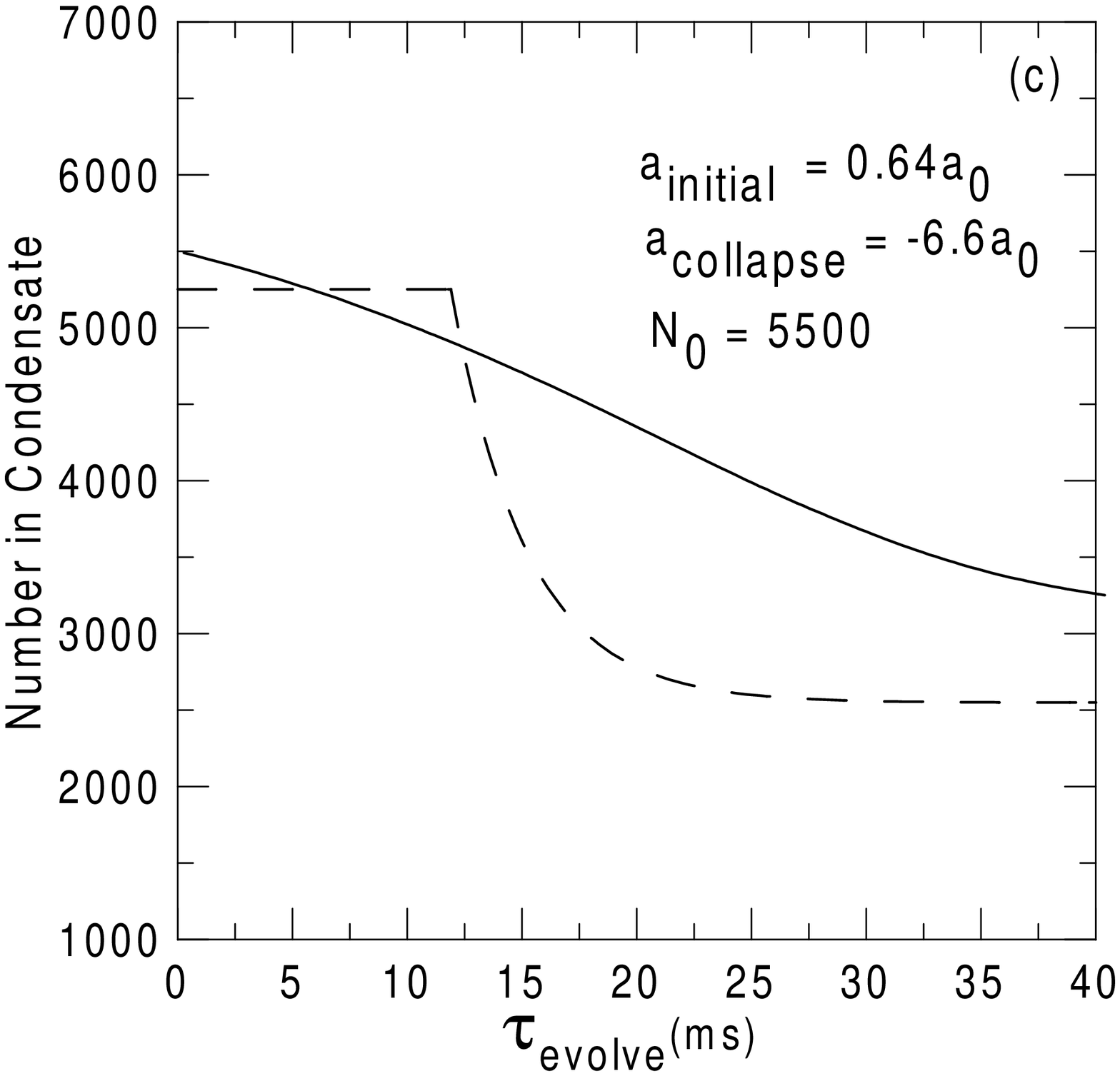}
\end{center}
 
\end{figure}
\vskip -1.5cm
\begin{figure}[!ht]
 
\begin{center}
\includegraphics[width=\linewidth]{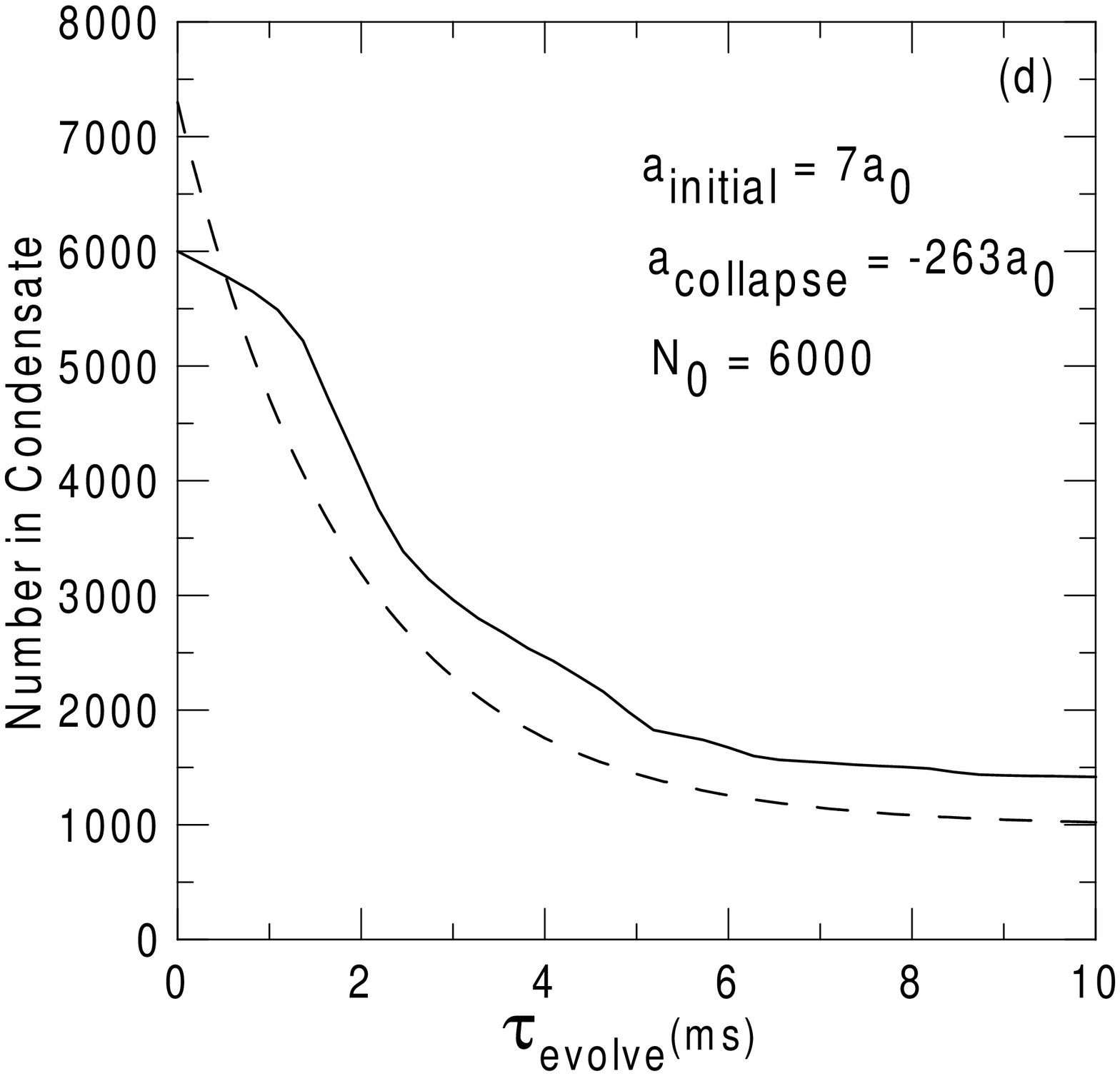}
\end{center}
 
\caption{More decay curves with $\xi = 2$ for (a) $a_{\mbox{initial}}=89a_0$,
$a_{\mbox{collapse}}=-15a_0$, and $N_0=6000,$ (b)
$a_{\mbox{initial}}=0.64a_0$, $a_{\mbox{collapse}}=-6.6a_0$, and
$N_0=14500$, (c) $a_{\mbox{initial}}=0.64a_0$,
$a_{\mbox{collapse}}=-6.6a_0$, and $N_0=5500$, and (d)
$a_{\mbox{initial}}=7a_0$, $a_{\mbox{collapse}}=-263a_0$, and $N_0=6000$.
Full line: present theory; dashed line: average over preliminary,
unanalyzed data using Eq. (\ref{dec}) \cite{don1}.}

\end{figure}

Donley et al.  fitted the decay in the  number of atoms
during particle
loss to a decay constant $\tau_{\mbox{decay}}$ via the formula
\begin{eqnarray} \label{dec}
N(\tau_{\mbox{evolve}})&=&N_{\mbox{remnant}}+(N_0-N_{\mbox{remnant}})
\nonumber \\ &\times&
e^{(t_{\mbox{collapse}}-\tau_{\mbox{evolve}})/\tau_{\mbox{decay}}}
\end{eqnarray} for $\tau_{\mbox{evolve}}>t_{\mbox{collapse}}$.
In Fig. 1 we also plot 
$N(\tau_{\mbox{evolve}})$ of Eq. (\ref{dec})  (dashed line)  for
$a_{\mbox{collapse}}=-263a_0, -30
a_0$ and $-6.7a_0$ with respective decay rates $\tau_{\mbox{decay}}
=$ 1.2 ms, 2.8 ms and 2.8 ms \cite{don1}. For wide variation of parameters
$a_{\mbox{initial}}, a_{\mbox{collapse}}, $ and $N_0$,
$\tau_{\mbox{decay}}$ varies approximately between 1 and 3.
The results of the present simulation (full line) agree well with the
average experimental result of Eq. (\ref{dec}) for three different
$a_{\mbox{collapse}}$ (dashed line) \cite{don1}.

Next we repeated our calculation for several other values for
$a_{\mbox{initial}}$, $a_{\mbox{collapse}}$, and $N_0$.
These results are plotted
in Fig. 3
for
(a) $a_{\mbox{initial}}=89a_0$, $a_{\mbox{collapse}}=-15a_0$, and
$N_0=6000,$
(b) $a_{\mbox{initial}}=0.64a_0$, $a_{\mbox{collapse}}=-6.6a_0$, and
$N_0=14500$,
(c) $a_{\mbox{initial}}=0.64a_0$, $a_{\mbox{collapse}}=-6.6a_0$, and
$N_0=5500$, and
(d) $a_{\mbox{initial}}=7a_0$, $a_{\mbox{collapse}}=-263a_0$, and
$N_0=6000$. The agreement of the result of simulation  with
unpublished, preliminary  unanalyzed
data
is good in all four cases reported in Fig. 3 \cite{don1}.

\begin{figure}[!ht]
 
\begin{center}
\includegraphics[width=\linewidth]{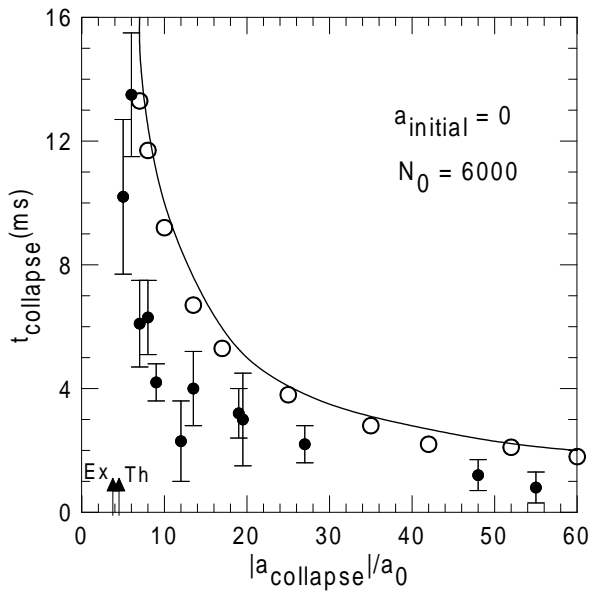}{1.}
\end{center}
 
\caption{The collapse time $t_{\mbox{collapse}}$ vs. $|a_{\mbox{collapse}}|/a_0$
for $a_{\mbox{initial}}=0$, $N_0=6000$ and $\xi=2$. Solid circle with
error bar: experiment \cite{ex4};  open circle: axially symmetric
mean-field model of Ref. \cite{th3}; arrows marked Th and Ex are
theoretical (4.49) \cite{sk3a,sk1} and experimental (3.75) \cite{ex3}
estimates of $|a_{\mbox{collapse}}|/a_0$, respectively, full line: present
theory.}

\end{figure}

The decay curves in Fig. 3 are  different, although they
have
certain general features which determine the decay constant
$\tau_{\mbox{decay}}$,
collapse time $t_{\mbox{collapse}}$, and number of atoms in the remnant.
Experimentally, the fraction of atoms that went
into the remnant decreased with $|a_{\mbox{collapse}}|$ and was $\sim
40\%$ for
$|a_{\mbox{collapse}}|<10a_0$ and was $\sim 10\%$
for $|a_{\mbox{collapse}}|>100a_0$.
Figures 1  and 3 also show this
behavior. The values of
$\tau_{\mbox{decay}}$ for plots in Figs. 3 (a) $-$ (d) are
1.5 ms, 2.4 ms, 3.3 ms, and 1.9 ms, respectively, lying in the range
$\sim 1 - 3$ ms \cite{don1}.  The general features  in the behavior of
remnant number
and collapse time are discussed in the following.

\begin{figure}[!ht]
 
\begin{center}
\includegraphics[width=\linewidth]{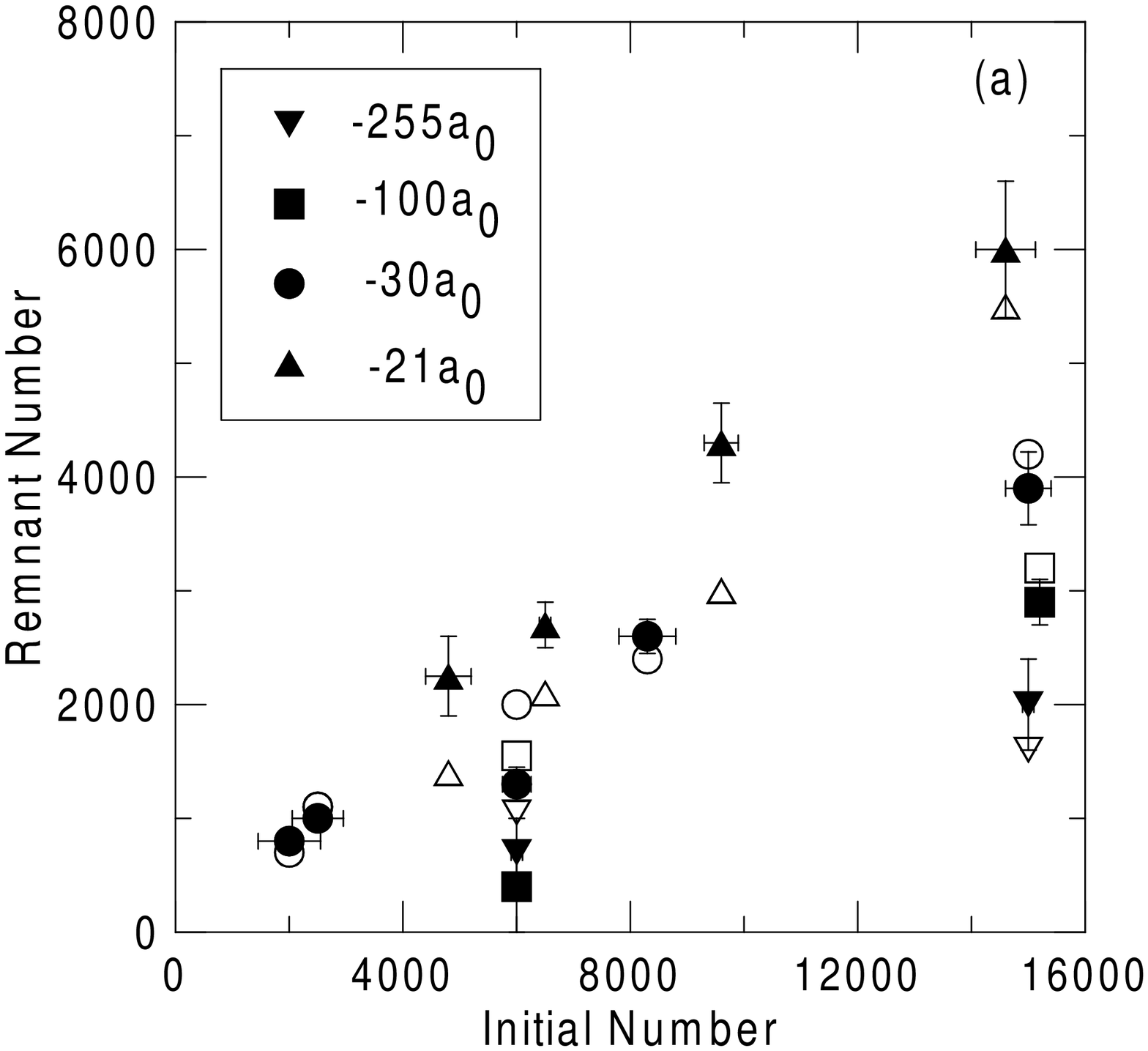}
\end{center}
 
\end{figure} 
\begin{figure}[!ht]
 
\begin{center}
\includegraphics[width=\linewidth]{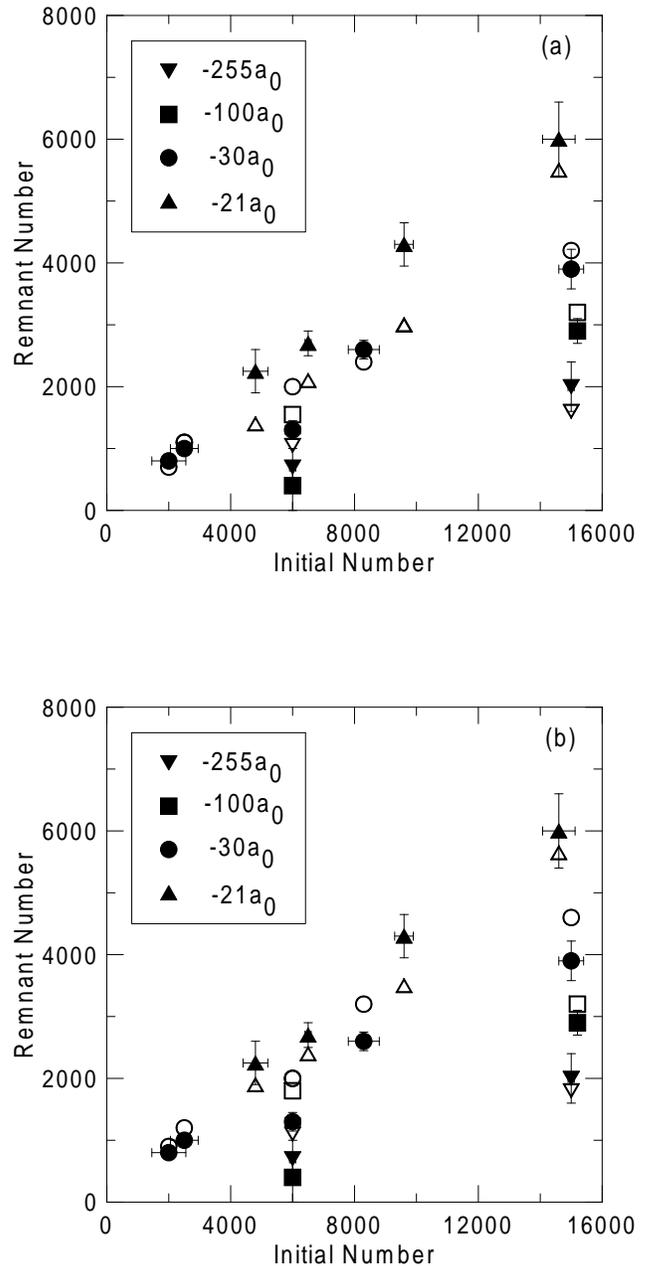}
\end{center}
 
\caption{Remnant number vs. initial number for $a_{\mbox{initial}}=7a_0$ and
different $a_{\mbox{collapse}}$ for (a) $\xi=3$ and (b) $\xi=2$. The
experimental results \cite{don2} with error bars are represented by solid
triangle, solid circle, solid square, and solid inverted triangle for
$a_{\mbox{collapse}}=-21a_0, -30a_0, -100a_0 $ and $-255a_0$. The
corresponding theoretical results are represented by open triangle, open
circle, open square, and open inverted triangle.}

\end{figure}

Donley et al. provided a quantitative measurement  of the variation of
collapse time $t_{\mbox{collapse}}$ with the final scattering length
$a_{\mbox{collapse}}$ for a given $a_{\mbox{initial}} =0$ and $N_0 =6000$.
We also calculated this variation using our model given by Eq. (\ref{d1}).
In our calculation we define $t_{\mbox{collapse}}$ as the time at which
the spikes (irregularities), as in Fig. 2, tend to appear in the wave
function. The results are plotted in Fig. 4 and compared with experimental
data \cite{ex4} as well as with another calculation using the mean-field
GP equation in an axially symmetric trap \cite{th3}. The agreement between
the two theoretical results is
very good. There is also qualitative
agreement between the experimental data on the one hand and the two
calculations on the other hand:  $t_{\mbox{collapse}}$ decreases with
$|a_{\mbox{collapse}}|/a_0$ starting from an infinite value at
$|a_{\mbox{collapse}}|=a_{\mbox{cr}}$ for a fixed $N_0$, that is
6000 in Fig. 4. For this $N_0$,
$a_{\mbox{cr}}$ is the
minimum value of $|a_{\mbox{collapse}}|$ that leads to the collapse and
explosion.
For a given $N_0$, a critical value of $n \equiv
n_{\mbox{cr}} $ for collapse can be defined
 via $n_{\mbox{cr}} \equiv  N_0
a_{\mbox{cr}}/l$. As
there is discrepancy between theoretical and experimental $n_{\mbox{cr}}$
for an axially symmetric trap \cite{sk3a,sk1}, the theoretical and
experimental $a_{\mbox{cr}}$ are also supposed to be different. The
experimental $k_{\mbox{cr}} \equiv   n_{\mbox{cr}}\lambda^{1/6} =  0.46 $
\cite{ex3} and the
theoretical
$k_{\mbox{cr}}  =  0.55 $ \cite{sk3a,sk1} for the axially
symmetric trap
used in the experiment at JILA with the asymmetry parameter
$\lambda = 0.389$.  The
theoretical $a_{\mbox{cr}}$ should be
larger than the experimental $a_{\mbox{cr}}$ in the same proportion.
This might imply that the theoretical $t_{\mbox{collapse}}$ should tend
to infinity for a slightly
larger value of $a_{\mbox{collapse}}$ as in Fig. 4.

\begin{figure}[ht]
 
\begin{center}
\includegraphics[width=\linewidth]{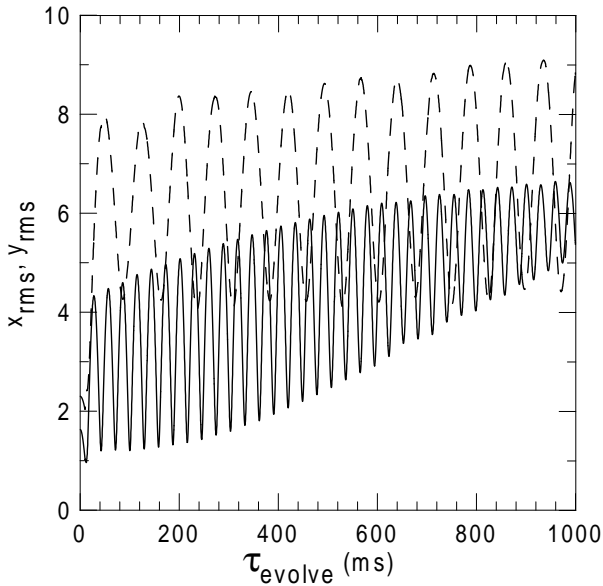}{1.}
\end{center}

\caption{The dimensionless rms sizes $x_{\mbox{rms}}$ (full line) and
$y_{\mbox{rms}}$ (dashed line) expressed in units of $l/\sqrt 2$ $
(l=26070$ \AA) after the jump in the scattering length of a BEC of 16000
$^{85}$Rb atoms from $a_{\mbox{initial}}=7a_0$ to
$a_{\mbox{collapse}}=-6.7a_0$ as functions of time $\tau_{\mbox{evolve}}$
for $\xi=2$.}

\end{figure}

Donley et al. measured the number of remnant atoms for
$a_{\mbox{initial}}=7a_0$ and different initial number $N_0$ and
$a_{\mbox{collapse}}$ and these results   \cite{don2} are plotted in
Figs. 5
(a) and (b)
and compared with numerical simulation
performed with   $\xi = 3$   and  2, respectively. The agreement is good
for most
cases
shown in this figure. For $N_0=6000$, there is some
discrepancy between theoretical and experimental remnant numbers.
The overall agreement is better in the case with
 $\xi = 2$ than with $\xi=3$. For $\xi =3$ the three-body recombination
loss-rate is larger and this leads to smaller remnant numbers compared to
the case with $\xi = 2$.
The theoretical  $N_{\mbox{cr}} $ for a fixed
negative $a_{\mbox{collapse}}$ is given by  $N_{\mbox{cr}}=
0.55l\lambda^{-1/6}/|a_{\mbox{collapse}} |$ \cite{sk3a,sk1}. For
$a_{\mbox{collapse}} = -255a_0, -100a_0, -30a_0,$ and $-21a_0$,
$N_{\mbox{cr}} = 124, 317, 1057, $ and 1510, respectively. Hence, from
Fig. 5 we find that  the number in the remnant
could be much larger than $N_{\mbox{cr}} $ for times on the order of tens
of milliseconds.  However, in our simulation such a remnant continues to
emit atoms at a much slower rate and for very large times
on the order of  seconds
the number of
atoms eventually tends towards $N_{\mbox{cr}}$.

Donley et al.  observed that the remnant condensate in all cases
oscillated in a highly excited collective state with approximate
frequencies $2\nu _{\mbox{axial}}$ and $2\nu _{\mbox{radial}}$ being
predominantly excited. The actual measured frequencies are 13.6(6) Hz and
33.4(3) Hz. To find if this behavior emerges from the present simulation
we plot in Fig. 6 sizes $x_{\mbox{rms}}$ and $y_{\mbox{rms}}$ vs. time for
the condensate after the jump in the scattering length to $-6.7a_0$ from
$7a_0$ for $N_0=16000 $. Excluding the first 20 ms when the remnant
condensate is being formed, we find a periodic oscillation in
$x_{\mbox{rms}}$ and $y_{\mbox{rms}}$ with frequencies 13.5 Hz and 34 Hz,
respectively, as observed in experiment.

\section{Discussion}

Though we have explained some aspects of the experiment at JILA, certain
detailed features have not been addressed in this study. Donley et al.
have classified the emitted atoms in three categories: burst, missing
(undetected) and jet atoms \cite{ex4}.  The jet atoms appear with much
lower energy solely in the radial direction possibly from the spikes in
the wave function when the collapse is suddenly interrupted during the
period of atom loss before the remnant is formed. Strangely enough the
emission of jet atoms are found not to possess axial symmetry always and
hence can not be properly treated in a axially symmetric model. Moreover a
clear-cut distinction between the burst and missing atoms emitted during the 
explosion seems to be difficult in the present model as the experiment
could not specify the properties (magnitude and direction of velocities)
of the missing atoms.  Also, because of the missing atoms it is difficult
to predict the energy distribution of the burst atoms during the explosion in
a mean-field analysis. Without proper identification of the missing atoms,
any energy distribution calculated using the present mean-field analysis
will yield the total energy of burst plus missing atoms.  A careful
analysis of the energy of the emitted atoms is required for explaining the
exclusive features and a detailed study of the wave function is needed for
this purpose. Such an analysis is beyond the scope of the present
investigation and would be a welcome future theoretical work. 
 
The success of the Crank-Nicholson algorithm in alternate directions as
used in this study depends on a proper discretization of the GP equation
in space and time. In this study we employed a two-dimensional lattice in
space of $600\times 150$ or 90000 points ($x\le 15, -30\le y \le 30$)
and a time step of
0.001. In the absence of collapse and recombination loss this
discretization leads to very precise results. The accuracy reduces in the
presence of the violent collapse and explosion simulated by
three-body recombination. By varying the space discretization grid
and time step we found that the estimated error in the present calculation
is less than $\sim 10 \%$ for time propagation upto few tens of
milliseconds.

There has been another attempt to use the mean-field GP equation in an
axially symmetric trap \cite{th3} to explain the experiment of Ref.
\cite{ex4}. There are certain differences between the analysis of Ref.
\cite{th3} and the present investigation.  According to the experiment of
Ref. \cite{ex4}, the burst atoms and missing atoms are components of
expelled atoms which lose contact with the central condensate that
eventually forms the remnant. Of these the burst atoms have energy much
less than the magnetic trap depth. Hence, though expelled from the central
condensate they continue trapped and oscillate with time. The wave
function of Eq. (\ref{d1}) only describes the central condensate. However,
in Ref. \cite{th3} the burst atoms are considered to be the peripheral
part (the spikes) of the central condensate and hence taken to be
described by the mean-field Eq. (\ref{d1}). The missing atoms are actually
parts of the expelled atoms that have disappeared from the trap
\cite{ex4}. In Ref. \cite{th3}, the missing atoms have been taken to be
 the only component of the emitted atoms. These are the main differences
between the point of view of the present analysis and that of Ref.
\cite{th3}.

The three-body loss rates of the two studies are also widely different.
Here we employ the three-body recombination loss-rate $K_3\simeq 9\times
10^{-25}$ cm$^6$/s for $a=-370a_0$ whereas in Ref. \cite{th3} the value
$K_3\sim 10^{-28}$ cm$^6$/s has been considered. The present rate is in
rough agreement with the experimental rate of Ref. \cite{k3} ($K_3\sim 4.2
\times 10^{-25}$ cm$^6$/s) and with the theoretical rate of Ref.
\cite{esry} ($K_3\sim 6.7 \times 10^{-25}$ cm$^6$/s) for the same value of
scattering length whereas that of Ref. \cite{th3} is orders of magnitude
smaller.  However, such a small three-body rate in Ref.  \cite{th3} has
led to a large residual condensate at large time that they have
interpreted as the sum of burst plus remnant.  The use of a large
three-body rate in this study has led to a much smaller residual central
condensate which has been identified as the remnant as in the experiment
at JILA \cite{ex4}.

However, it is assuring to see that the $t_{\mbox{collapse}}$ vs.
$|a_{\mbox{collapse}}|/a_0$ curve of the two models in Fig. 4 agrees with
each other. The present calculation in Fig. 4 was performed with a nonzero
loss rate $K_3$, whereas that in Ref. \cite{th3} was performed by setting
$K_3=0$. We find that $K_3$ plays an insignificant role in this
calculation at small times. Hence, the two computer routines lead to the
same result in the absence of recombination loss before the beginning of
the explosion.

\section{Conclusion}

In conclusion, we have employed a numerical simulation based on the
accurate
solution \cite{sk1}
of the mean-field Gross-Pitaevskii equation with a cylindrical
trap to study the dynamics of the collapse and explosion as observed in the
recent experiment at JILA \cite{ex4}. In the GP equation we include a
quintic
three-body nonlinear recombination loss term that accounts for the decay
of the strongly attractive condensate.  The results of the present
simulation accounts for some aspects of the experiment.

In the experiment a strongly attractive $^{85}$Rb condensate was
prepared by ramping the scattering length to a large negative value and
the subsequent decay of the collapsing and exploding
condensate was measured. We have been able to
understand the following features of this dynamics from the present
numerical simulation: (1) The condensate undergoes  collapse
and explosion and finally stabilizes to a remnant condensate containing
about $\sim 10\%$ (for $|a_{\mbox{collapse}}|>100a_0$ ) to   $40\%$ (for
$|a_{\mbox{collapse}}|<10a_0$)
of
initial number of atoms $N_0$ at large times. This percentage
is  independent of
 $N_0$ and the ramped scattering length $a_{\mbox{collapse}}.$ The number
in the remnant condensate can be much larger than the critical number for
collapse $N_{\mbox{cr}}$ for the same atomic interaction for experimental
times on the order of tens of milliseconds.    (2) Both in the
experiment and our simulation the remnant condensate executes radial and
axial oscillations in a highly excited collective state for a long time
with frequencies $2\nu_{\mbox{radial}}$ and $2\nu_{\mbox{axial}}.$ (3)
After the sudden change in the scattering length to a large negative
value, the condensate needs an interval of time $t_ {\mbox{collapse}}$
before it experiences loss via explosion.  Consequently, the decay starts
after the interval of time $t_ {\mbox{collapse}}$. (4) The number of atoms
in the condensate decays exponentially with a decay constant
$\tau_{\mbox{decay}}$ of few milliseconds ($\sim 1-3$ ms).
 
To conclude,
a large part of the Bosenova  experiment on $^{85}$Rb atoms
at JILA
\cite{ex4}, specially the detailed behavior of the remnant,
can be understood by introducing  the rather conventional
three-body recombination loss in the standard mean-field GP equation,
  with a loss rate compatible with other studies \cite{k3,esry}. The study
of the detailed behavior of the burst and missing atoms and the formation
of the jet in such a mean-field theory seems to be more complicated
technically, nevertheless viable, and would be a subject of future
investigation.

\acknowledgments

The author extends sincere thanks to Dr. E. A. Donley for additional
unpublished data \cite{don1,don2}  of the experiment at JILA \cite{ex4}.
The work is supported in part by the CNPq and FAPESP
of Brazil.

\end{document}